\documentclass[
  superscriptaddress,
  amsmath,amssymb,
  aps,
  reprint,
  prb,
]{revtex4-1}

\usepackage{graphicx}  
\usepackage{siunitx}
\usepackage{caption}
\usepackage{subcaption}  
\usepackage{dcolumn}
\usepackage{bm}
\usepackage{physics}
\usepackage{comment}
\usepackage[normalem]{ulem}
\usepackage[colorlinks=true, citecolor=blue, linkcolor=blue, urlcolor=blue]{hyperref}
\usepackage[capitalize, nameinlink]{cleveref}

\newcommand{\figreflet}[2]{\hyperref[#1]{Figure~\ref*{#1}#2}}
\newcommand{\trise}{$t_\mathrm{rise}$ }
\newcommand{\tfall}{$t_\mathrm{fall}$ }
\newcommand{\tdelta}{$\Delta t$ }

\begin{document}

\title{Dual-Trigger of Series Nanowire Detector for Event-Based Photon Number Assignment}
\date{\today}

\author{Stefanie Grotowski}
\email{stefanie.grotowski@tum.de}
\affiliation{Walter Schottky Institut, Technische Universität München, 85748 Garching, Germany}
\affiliation{TUM School of Natural Sciences, Technische Universität München, 85748 Garching, Germany}
\affiliation{Munich Quantum Instruments GmbH, 85748 Garching, Germany}

\author{Daniel Landes}
\affiliation{Walter Schottky Institut, Technische Universität München, 85748 Garching, Germany}
\affiliation{TUM School of Natural Sciences, Technische Universität München, 85748 Garching, Germany}
\affiliation{Munich Quantum Instruments GmbH, 85748 Garching, Germany}

\author{Fabian Wietschorke}
\affiliation{Walter Schottky Institut, Technische Universität München, 85748 Garching, Germany}
\affiliation{TUM School of Computation, Information and Technology, Technische Universität München, 80333 München, Germany}

\author{Paul Pucknus}
\affiliation{Walter Schottky Institut, Technische Universität München, 85748 Garching, Germany}
\affiliation{TUM School of Natural Sciences, Technische Universität München, 85748 Garching, Germany}
\affiliation{Munich Quantum Instruments GmbH, 85748 Garching, Germany}

\author{Rasmus Flaschmann}
\affiliation{Munich Quantum Instruments GmbH, 85748 Garching, Germany}

\author{Torsten Langer}
\affiliation{PicoQuant GmbH, 12489 Berlin, Germany}

\author{Torsten Krause}
\affiliation{PicoQuant GmbH, 12489 Berlin, Germany}

\author{Kai Müller}
\affiliation{Walter Schottky Institut, Technische Universität München, 85748 Garching, Germany}
\affiliation{TUM School of Computation, Information and Technology, Technische Universität München, 80333 München, Germany}
\affiliation{Munich Center of Quantum Science and Technology (MCQST), 80799 München, Germany}

\author{Jonathan J. Finley}
\affiliation{Walter Schottky Institut, Technische Universität München, 85748 Garching, Germany}
\affiliation{TUM School of Natural Sciences, Technische Universität München, 85748 Garching, Germany}
\affiliation{Munich Center of Quantum Science and Technology (MCQST), 80799 München, Germany}

\begin{abstract}
    Photon-number-resolving (PNR) detectors are essential components of photonic quantum technologies.
    However, conventional single-channel edge-triggered readout struggles to resolve the photon number $n$ in real time or to characterize how timing jitter depends on $n$.
    In this work, we use a dual-trigger method on a three-pixel series-connected superconducting nanowire single-photon detector (SNSPD) that triggers on both the rising and falling edges of the detection pulse. 
    By doing so, we preserve the precise arrival time of the detection event while mapping the photon number onto the time interval between the rising and falling edges, allowing clear separation of the events.
    Using this technique, we assign each detection event to $n = 1, 2,$ or 3 photons with $99\,\%$ posterior confidence across all three classes.
    The timing jitter decreases as $n$ increases, reaching values below $\SI{41}{\pico \second}$ for $n \geq 2$. 
    Comparing edge-triggering with constant-fraction discrimination (CFD) for arrival-time extraction, we find CFD yields lower jitter for single-photon events and a nearly constant mean arrival time.
    Altogether, our results establish dual-triggering as a robust, low-latency readout scheme for PNR detectors, while revealing a photon-number dependence of the timing jitter relevant to timing precision achievable in heralded photonic quantum applications.
\end{abstract}

\maketitle

\section{Introduction}
Photon-number-resolving (PNR) detectors are of great interest for many applications in photonic quantum technologies, finding application in quantum communication \cite{Becerra2015}, photonic quantum computing \cite{Nerenberg2025, OBrien2016} and random number generation \cite{Applegate2015}. 
Early approaches to realize PNR were based on time multiplexing, where the light is distributed into multiple paths with different travel times and the photons are detected with an ordinary single-photon detector \cite{Fitch2003}. 
By now, many concepts have been proposed to achieve PNR. 
Multiple, individual superconducting nanowire single-photon detectors (SNSPDs) can be used and arranged in an interleaved manner \cite{Dauler2007, Resta2023}. 
Each pixel has its own RF line and is individually biased. 
This allows for combining PNR functionality with a large detector area while maintaining low jitter due to short pixel nanowires.
The photon number can be determined from coincidence events, which are counted on individual channels of the time-tagging device. 
This comes with the downside that at the cryogenic level a readout line is required for each pixel, limiting the approach's scalability. 
To reduce the number of lines, multiple SNSPDs can be biased from a single cable, with the detectors operated in parallel \cite{Stasi2024}.
To allow a homogeneous current distribution and avoid switching of the entire device when a photon is absorbed, advanced circuits using resistors \cite{Huang2018} or resistors in series with inductors \cite{Perrenoud2021} were developed. 
Another approach that brings the same advantage of a shared bias and readout line is the series nanowire detector (SND). 
The concept of an SND was first proposed by Jahanmirinejad \textit{et al.} \cite{Jahanmirinejad2012} in 2012.
In this architecture, the photon number is encoded in the amplitude of the output signal. 
This device consists of $N$ individual SNSPDs connected in series, each of them shunted with a parallel resistor $R_\mathrm{p}$, where the output amplitude scales with the number of clicking pixels. 
The SND is biased with a constant current usually provided by a voltage source and a pre-resistor. 
Once a photon is absorbed in one of the pixels, the current is redirected into the parallel shunt, minimizing the current reduction in the unfired pixels, which remain sensitive \cite{Tao2020}. 
SNDs preserve all the advantages of SNSPDs in terms of simplicity, sensitivity, timing resolution and speed. 
Previous studies showed the extraction of photon-number over the distribution as a statistical property \cite{Cahall2017}, where especially, in overlap regions the photon number cannot be clearly determined. 
Photon-number-dependent properties, such as timing jitter are resolved by post-processing methods \cite{Hao2024}. 
The main challenge in measuring timing jitter at the 1-, 2- and 3-photon levels is that isolating individual photon levels requires adjusting both the mean photon number and the trigger threshold — yet the timing jitter itself depends on both of these quantities \cite{Sidorova2025, Mueller2023}. \\

In this work, we present an event-based, high-fidelity photon-number assignment method. 
We show how photon levels can be separated using a dual-trigger method that triggers on the rising and falling edge of the output signal. 
We start by analyzing the photon-number-dependent waveforms and identifying the desired trigger levels on both the rising and falling edge. 
This allows us to map the photon-number encoding from the pulse amplitude onto a measurable time delay, while maintaining the arrival timestamp.
This provides a much more pronounced separation, less overlap in the $n$-photon distribution and greater distinguishability. 
Moreover, we compare two trigger mechanisms, the constant-fraction discrimination (CFD) and edge-trigger mode.
We can assign photon numbers with a certainty of $99\,\%$ while having a leakage rate of unassigned photons of $0.01\,\%$.
Lastly, we extract the photon-number-dependent timing jitter of the device. 

\section{Methods}
The test device is an SND made of three pixels connected in series, each shunted with a parallel resistor, allowing PNR and is provided by \textit{Munich Quantum Instruments}. 
The device is measured inside an ADR cryostat (\textit{Kiutra}) at a base temperature of \SI{2.3}{\kelvin}. 
The maximum current at which the device can be operated is \SI{10}{\micro \ampere}. 
We bias the detector at a constant current of \SI{8}{\micro \ampere}.
After photon absorption, the RF output is separated from the DC circuit via a bias tee. 
The signal is first amplified at \SI{3}{\kelvin} using a cryogenic amplifier (\textit{Cosmic Microwave Technologies, CITLF3}) and then again at room temperature (\textit{RF Bay, LNA2000}). 
We record the signal either with a fast oscilloscope (\textit{Tektronix, MSO64}) or Time-Correlated Single Photon Counting (TCSPC) device (\textit{PicoQuant, PicoHarp 330}). 
The sample is illuminated with a \SI{780}{\nano \meter} pulsed laser. 
The pulse duration is \SI{2}{\pico \second} and the repetition rate is set to \SI{8}{\mega\hertz} using an acousto-optic modulator (AOM). 
Furthermore, we can vary the mean photon number per pulse by a variable optical attenuator (VOA). 

\section{Results}
\begin{figure}[t]
\includegraphics[width=\columnwidth]{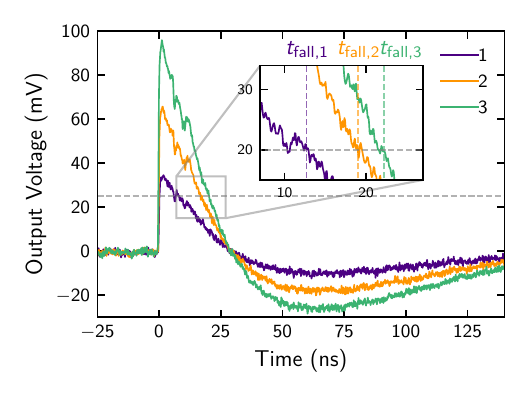}
    \caption{Waveform schematic of how the trigger levels of the rising and falling edge are chosen to achieve separation in the 2D heatmap of the photon number.}
    \label{fig:PNR_waveforms}
\end{figure}

In this section, we perform an in-depth study of the PNR performance of the SND detector and relate it to the corresponding pulse amplitudes.
We isolate the photon levels by a combination of mean photon number and threshold.
\Cref{fig:PNR_waveforms} shows the waveforms for the individual photon levels using a fast oscilloscope.
The photon number is encoded in the pulse amplitude, which increases linearly with photon number.
The 1-photon, 2-photon and 3-photon events have a mean pulse amplitude of \SI{35}{\milli \volt}, \SI{66}{\milli \volt} and \SI{96}{\milli \volt}, respectively. 
We see that all photon levels have a well-defined rising edge. 
The rise time is defined as the time interval between 1/$e$ and 90\% of the maximum amplitude. 
It stays constant with $n$ and has a mean value of $\tau_\mathrm{rise} = \SI{658}{\pico \second}$. 
Together with the increase in pulse height with photon number, the slew rate increases. 
The mean fall time is $\tau_\mathrm{fall} = \SI{14.3}{\nano \second}$ and is defined as the time constant of the pulse decay between 90\% and 1/$e$ of the maximum amplitude. 
These time constants are independent of the number of pixels clicking, as the decay constant of the SND is defined by the total inductance of the device $\tau \propto N \cdot L_\mathrm{k}/R_\mathrm{L}$ where $N$ denotes the number of pixels and $R_\mathrm{L}$ the load resistance \cite{Jahanmirinejad2012}. 
The detailed statistical evaluation can be found in \Cref{sec:pnr_pp}. 
We find that the amplifier causes an undershoot that scales with the maximum voltage transient amplitude.
After approximately \SI{140}{\nano \second}, the signal returns to its baseline. 
Our measurement is performed at a repetition rate of \SI{8}{\mega \hertz}, corresponding to a pulse separation of \SI{125}{\nano \second}.
This could lead to time-walk effects for pulses arriving before the pulse has fully returned to its baseline \cite{Mueller2023}. 
The PNR capabilities at this repetition rate are not compromised by the undershoot as the falling edge shape is preserved. \\

\begin{figure*}[t]
    \centering
    \includegraphics[width=\textwidth]{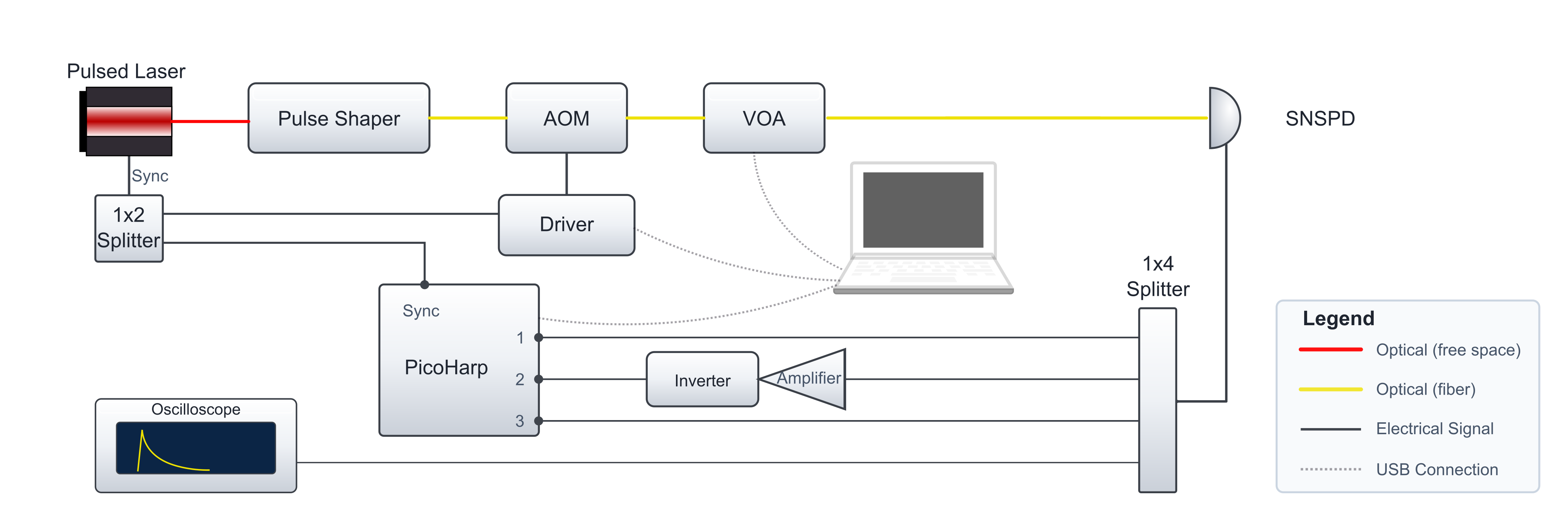}
    \caption{Schematic overview of used the measurement setup. The excitation light is guided to the SNSPDs through a pulse shaper, an AOM and VOA to control the pulse duration, repetition rate and attenuation, respectively. The SNSPD output signal is split into four channels, with three connected to the PicoHarp and one serving as a reference on the oscilloscope. The signal at channel (ii) is amplified and inverted to be compatible with the CFD mode. }
    \label{fig:schematic}
\end{figure*}
Other works show an intrinsic photon-number-resolution by recording the events with two trigger levels on the rising edge \cite{Los2024} to make use of intrinsic photon number resolution capability of SNSPDs \cite{Schapeler2024}. 
In this work, we set two trigger levels on the rising and falling edge, respectively, and take advantage of the fact that the pulse amplitude varies with photon number while exhibiting a constant decay time for all photon numbers. 
As the pulse height scales with photon number, it takes longer for the pulse to decay to a given threshold level.
The separation is mapped from the pulse amplitude to a measurable time above threshold \cite{Tiedau2020}. 
The inset shows the time delays on the falling edge, which are denoted by $t_\mathrm{fall, 1}$, $t_\mathrm{fall, 2}$, and $t_\mathrm{fall, 3}$, respectively. 
As shown in \Cref{fig:PNR_waveforms}, we find that $t_\mathrm{fall, 1} = \SI{12.7}{\nano \second}$, $t_\mathrm{fall, 2} = \SI{19}{\nano \second}$ and $t_\mathrm{fall, 3} = \SI{22.2}{\nano \second}$ for a trigger level of \SI{20}{\milli \volt}. 
Hence, triggering on the falling edge allows us to clearly distinguish the 1, 2 and 3 photon level while maintaining the arrival time by the rising edge-trigger. 
The grouping can be performed by assigning fixed delay time windows to each photon number. 
The boundaries of these time intervals can be determined by analysis of an event distribution, as will be shown later. 
This provides a higher certainty, compared to results from previous works \cite{Schapeler2024} which provided a probabilistic photon number distribution. 

\begin{figure*}[t]
    \centering
    \includegraphics[width=\textwidth]{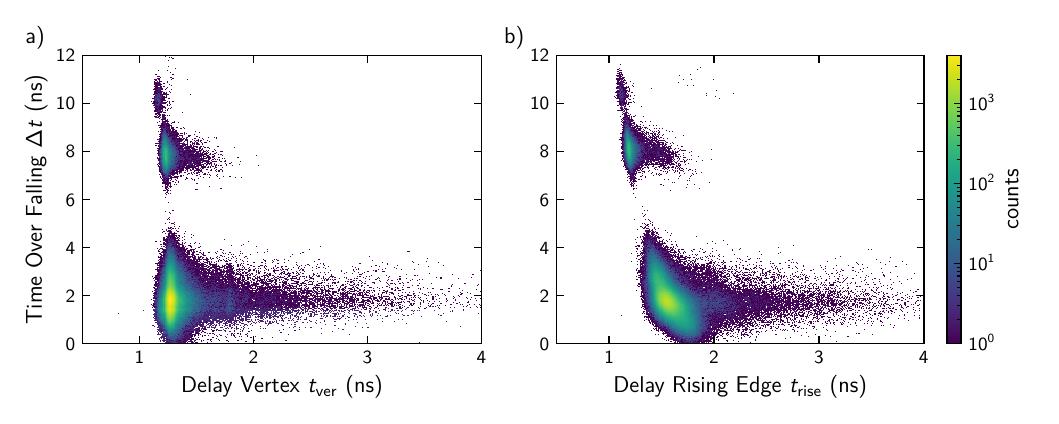}
    \caption{2D heatmap of the delay of the rising edge and falling edge relative to the sync signal. a) CFD mode and b) Edge mode on rising edge. A clear separation of the photon levels in island shapes is visible. }
    \label{fig:heatmap}
\end{figure*}

In our experiment, we first split the signal into four channels (\textit{Mini Circuits, ZFRSC-4-842-S}), allowing us to independently control the trigger mode and trigger level for each channel.
Regarding the two trigger mechanisms, the edge-trigger provides the timestamp when the voltage transient crosses a given threshold. 
In contrast, the CFD discriminates pulses based on their amplitude, similar to an edge-trigger. 
However, the timestamp is assigned at the vertex of the respective pulses. Consequently, the resulting time tag is defined by an effective constant fraction of 1 (i.e. \SI{100}{\percent}) of the maximum amplitude, which is invariant to changes in the pulse amplitude.
The measurement utilizes (i) an edge-trigger on the rising edge at \SI{15}{\milli \volt}, (ii) a CFD trigger with a discriminator at \SI{-85}{\milli \volt}, (iii) edge-trigger on the falling edge at \SI{25}{\milli \volt} and (iv) an oscilloscope reference.
The CFD input signal at channel (ii) is first amplified to ensure sufficient amplitude and then inverted for compatibility with the TCSPC device using a SIA400, a component featuring a high-pass filter. 
An overview of the measurement setup is provided in the figure at the top of this paragraph. 
In \figreflet{fig:heatmap}{a} we show the two-dimensional event map.
The photon events are shown on the x-axis by their arrival times relative to the sync signal of the pulsed laser. 
Depending on the trigger mechanism used this timestamp is defined as $t_\mathrm{ver}$ as the vertex of the pulse, or $t_\mathrm{rise}$ the timestamp given by the rising edge for CFD and edge-trigger, respectively. 
On the y-axis, the detection events are separated by difference in delay times of the falling edge \tfall ($t_\mathrm{fall, 1}$, $t_\mathrm{fall, 2}$, $t_\mathrm{fall, 3}$) to the arrival time, which we denote as time over falling $\Delta t$.
Since both timestamps are referenced to the same trigger pulse, computing their difference removes the common‑mode jitter contribution from the sync signal itself.
Furthermore, for simplicity, we offset the y-axis data so that it begins at zero.
We identify three islands, which correspond to each photon level. 
The island with the lowest \tdelta $<$ \SI{6}{\nano \second} corresponds to the single-photon event. 
Since the single-photon event has the lowest amplitude, the delay between the two timestamps should be minimal.
With increasing photon number \tdelta increases to \SI{6}{\nano \second} $<$ \tdelta $<$ \SI{9.5}{\nano \second} for the 2-photon event. 
With higher amplitude but constant fall time the time span between the rising and falling event is expected to be larger than for the single-photon event. 
Lastly, the delay window of \SI{9.5}{\nano \second} $<$ \tdelta $<$ \SI{12}{\nano \second} allows us to identify the 3-photon event. 
These results also match well with the waveforms shown in \Cref{fig:PNR_waveforms} showing the clear separation between $t_\mathrm{fall, 1}$, $t_\mathrm{fall, 2}$ and $t_\mathrm{fall, 3}$.
\figreflet{fig:heatmap}{a} shows the data of channels (ii) and (iii), where the rising edge is recorded in CFD mode. 
Comparing the three islands, we see that the 1-photon event has the largest spread and tail towards larger $t_\mathrm{ver}$. 
There are multiple reasons that explain this observation. 
Despite the high uniformity of the fabrication process, the pixels are not expected to be perfectly identical. 
In particular, a variation of the shunt resistors of each branch can lead to a slight variation in the output voltage transient and influence the pulse height. 
Therefore, each clicking pixel would have its own signature, contributing to the total 1-photon jitter. 
Moreover, the position of the firing pixel adds a geometric contribution to the timing jitter. 
As they are connected in series, the signal propagation time depends on the location of the absorption site. 
Additionally, potential detection events in the taper region of the detector can lead to a tail towards larger delay times, which is why the distribution is usually fitted using an exponentially modified Gaussian (EMG) function \cite{Sidorova2017, Korzh2020}. 

The same dataset recorded with an edge-trigger on the rising edge from channel (i) is presented in \figreflet{fig:heatmap}{b}. 
One major difference is visible in the lowest island for the single-photon event, which shows a much larger spread across $t_\mathrm{rise}$. 
We expect the main contribution to come from the threshold setting, which influences the timing jitter. 
The jitter is minimized when setting the threshold close to \SI{50}{\percent} of the maximum pulse height, where the slope is maximized. 
However, in this measurement configuration, the threshold is set to be suitable for all photon numbers, and not specifically optimized for low single-photon jitter. 
Additionally, a time-walk contribution of the timing jitter arises from variations in the pulse amplitude due to the amplifier's undershoot behavior. 
At \SI{8}{\mega\hertz}, the \SI{125}{ns} pulse separation is too short for full recovery to baseline.
Depending on the photon number of the previous detection event, the baseline is at a different voltage, causing a variation in pulse height, which is reflected in a delay in the threshold crossing \cite{Mueller2023}. 
Moreover, we observe a shift of the mean delay value of the rising edge to shorter arrival times with increasing photon number, which can also be attributed to the triggering mechanism. 
With similar rise time and increased pulse height for higher photon numbers, we expect these to cross the threshold after shorter time delays.
Comparing these two trigger mechanisms, we consider the CFD mode to be advantageous for quantum applications. 
Especially for communication protocols, the bin size and timing jitter are essential, as they determine the data transmission rate \cite{Gouzien2018}. 
Because the mean arrival time shifts less, smaller time bins can be chosen.
In addition, we observe that the distribution of the arrival times in the x-direction is broader and less confined.
The effects mentioned above can be prevented by using the CFD mode, where the timestamp is determined by the vertex of the pulses. 
In our case, this is two-fold beneficial: First, it can provide a timestamp equally precise for all photon levels. 
Moreover, it compensates for small deviations in pulse amplitude at the 1-photon level. 
The beneficial effects of the CFD are expected to be even larger when higher repetition rates than \SI{8}{\mega \hertz} are chosen since then negative effects from the time-walk effects are anticipated to be larger. \\

In this section, we want to calculate the quality and confidence of photon-number assignment by $\Delta t$. 
To do so, we collapse the data along $t_\mathrm{ver}$, as this information is not used for the assignment, and model the data. 
Previous works showed that the arrival times are most accurately modeled by an exponentially modified Gaussian (EMG) distribution \cite{Schapeler2026a, Sidorova2017}. 
Consequently, a similar distribution is expected for $\Delta t$, which we fit simultaneously using a sum of three EMG distributions with Poisson (1/$\sqrt{N}$) weighting to account for the $\sim$4 orders of magnitude spanned by the peak populations.
\begin{figure}[h]
\includegraphics[width=\columnwidth]{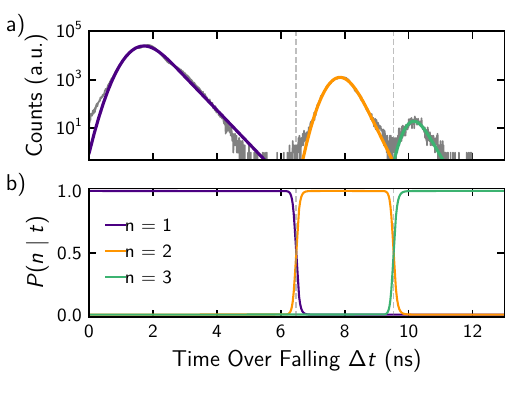}
    \caption{a) Data collapsed along the x-axis to resolve the photon-number encoding given by $t_\mathrm{fall}$. The data is modeled with EMG functions. b) Calculated weighted probabilities for the photon number $n$ at a given time delay $t$. }
    \label{fig:purity}
\end{figure}

\begin{figure}[h]
\includegraphics[width=\columnwidth]{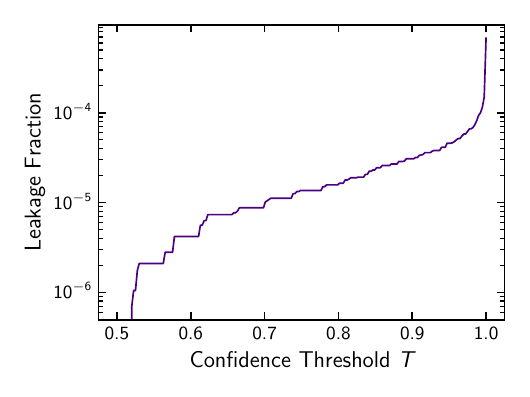}
    \caption{Te fraction of measured events with $P(n^\star\mid t) $<$ T$, as a function of the confidence threshold $T$. At $T=0.99$ the leakage fraction remains below \SI{0.01}{\percent}.}
    \label{fig:leakage}
\end{figure}

The result is depicted in \figreflet{fig:purity}{a} and we find good agreement between the data and the model. 
We now want to calculate the probabilities with which the photon numbers are assigned. 
Our excitation comes from a pulsed laser source of Poissonian nature, which is defined by a mean photon number $\mu$. 
We can calculate the detected mean photon number $\mu_\mathrm{det}$ by evaluating the zero-click events, similar to Schapeler \textit{et al.} \cite{Schapeler2024}. 
\begin{equation}
    \mu_\mathrm{det} = - \operatorname{ln}\left(1 - \ \frac{r_\mathrm{count}}{r_\mathrm{rep}}\right) 
\end{equation}
From this we obtain a mean photon number at the detector of $\mu_\mathrm{det} = 0.15$, which intrinsically accounts for all optical losses along the path from the laser source to the detector, including the fiber interconnects and the free-space coupling to the sample.
The EMG function $g_n(t)$ describes how likely it is that a detection event of photon number $n$ will arrive at delay $t$ and is given by the following equation 
\begin{equation}
\begin{split}
g_n(t) = \frac{1}{2\tau_n}
&\exp\!\left( \frac{\sigma_n^{2}}{2\tau_n^{2}} - \frac{t-\mu_n}{\tau_n} \right) \times \\
&\left[1 -  \operatorname{erf}\!\left( \frac{1}{\sqrt{2}}
\left( \frac{\sigma_n}{\tau_n} - \frac{t-\mu_n}{\sigma_n} \right) \right) \right].
\end{split}
\end{equation}
Here, $\mu_n$ describes the mean of the gaussian, $\sigma_n$ the width and $\tau_n$ the exponential tail. 
The mean photon number determined from the area ratios of the model yields a slightly lower value of $\mu = 0.12$. 
Due to the free space illumination scheme, we most likely do not have an ideal click distribution of the three-pixel device, leading to an underestimation of the 2- and 3-photon events and hence an underestimated $\mu$. 
Furthermore, we need to consider that our histogram contains only the recorded detection events. 
However, as mentioned above the underlying click-number distribution at these low photon numbers is dominated by the (unrecorded) zero-click events, which requires renormalizing the weights.
In order to describe the population in the histogram correctly, we add a weighting factor to each probability and renormalize, using the robust, model-independent $\mu_\mathrm{det}$. 
To account for the click statistics, the probability $P(n \mid \mu_\mathrm{det})$ is given by the area of the $n$-photon distribution: 
\begin{equation}
    \tilde{w}_n = \frac{P(n \mid \mu_\mathrm{det}) }{\sum_{n=1}^{3} P(n \mid \mu_\mathrm{det}) }.
\end{equation}
Finally, we can use Bayes' Theorem to determine the conditional probability, when a detection event arrives at time delay $t$, that it is of photon number $n$: 
\begin{equation}
P(n \mid t) = \frac{\tilde{w}_n\, g_n(t)}
{\displaystyle\sum_{j=1}^{3} \tilde{w}_j\, g_j(t)}
\end{equation}
The probability arising from our model is shown in \figreflet{fig:purity}{b}. 
We can identify three distinct regions in which a single photon number dominates. 
Having assigned the probabilities, we next extract different quantification metrics to compare the results. 
We start with the boundary which is set at the time delay where the probabilities of two neighboring photon events are equal. 
\begin{equation}
    P(i \mid b_{ij}) = P(j \mid b_{ij}) .
\end{equation}
The boundaries are $b_{12} = \SI{6.48}{\nano \second}$ and $b_{23} = \SI{9.53}{\nano \second}$ and can be used for event-based assignment.
To bound the effect of the prior, we recompute the weighting for both the area-ratio derived $\mu = 0.12$, which yields a second set of weights $\tilde{w}_n$. 
This shifts the prior toward lower click numbers but, owing to the clear separation of the peaks, does not move the boundaries.
To assess the quality of the assignment, we quantify the leakage fraction: the fraction of events whose maximum posterior probability falls below a chosen confidence threshold $T$. 
We map the measured data directly onto the distribution by using \Cref{eq:mapping}. 
\begin{equation}
y_n(t) = y(t)\, P(n \mid t)
\label{eq:mapping}
\end{equation}
From this direct mapping, an event at delay $t$ contributes to the leakage fraction at threshold $T$ if its maximum posterior probability, $P(n^\star \mid t)$ with $n^\star = \arg\max_n P(n\mid t)$, falls below $T$. 
\Cref{fig:leakage} shows the resulting leakage fraction as a function of $T$. 
We find that even at $T = 0.99$, the leakage fraction remains below $0.01\,\%$ and that it remains unchanged for the area-ratio derived $\mu$. 
Consequently, the assignment confidence is expected to increase further for lower mean photon numbers. 
If we move towards higher mean photon numbers, we approach the partially resolved regime, increasing the overlap between the 2- and 3-photon events. 
However, this can be further engineered by adapting the resistive circuit and increasing the overall amplitude, which would result in a larger spacing in $\Delta t$. 

In summary, it has been shown that the photon number assignment is determined by the pulse shape, mapping the amplitude onto a time delay. 
\\

In the last section we want to make use of this method and unlock the quantification of photon-number-dependent timing jitter. 
This metric is strongly threshold and count-rate-dependent \cite{Sidorova2025, Mueller2023}. 
However, we are able to separate the events without changing either of the two parameters to isolate the events.
We integrate all events assigned to one photon-number by \tdelta along our determined photon-level-boundaries $b_{12}$ and $b_{23}$. 
The distributions are fitted with an EMG function, where the FWHM of the distribution is referred to as timing jitter.
\begin{figure}[h]
\includegraphics[width=\columnwidth]{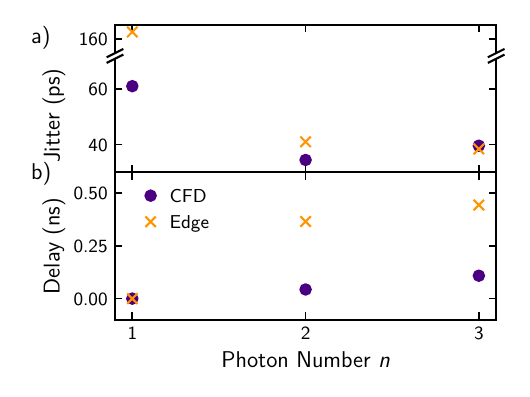}
    \caption{Extracted photon-number-dependent a) timing jitter and b) mean delay recorded with edge and CFD mode.}
    \label{fig:jitter}
\end{figure}
\figreflet{fig:jitter}{a} shows the timing jitter as a function of photon number $n$. 
We expect the timing jitter to decrease as the photon number increases due to a higher slew rate \cite{Flaschmann2023}, as discussed for \Cref{fig:PNR_waveforms} and an improving signal-to-noise ratio. 
Moreover, we observe a substantial difference in single-photon jitter between the two trigger modes, with the edge-trigger reaching \SI{161}{\pico\second} compared to only \SI{60}{\pico\second} for CFD.
This is similar to the discussion regarding \Cref{fig:heatmap}, where we attribute this to unoptimized threshold settings. 
Because of the working principle of the edge trigger, it allows only one threshold to be set for all photon levels. 
As a consequence, only an unoptimized threshold for the single-photon event can be used. 
In comparison, using the CFD this effect is mitigated, because the timestamp is given by the vertex of each pulse, independent on the pulse amplitude. 
Additionally, using the CFD mode is less sensitive to small variations in the pulse amplitude and consequently could contribute to a lower jitter value. 
For the higher photon numbers $n \geq 2$ we obtain timing jitter below \SI{41}{\pico \second} for all configurations. 

We continue with comparing the mean arrival times $t_\mathrm{ver}$ and \trise of the detection events relative to the sync signal in \figreflet{fig:jitter}{b}. 
For the edge-trigger we see a difference of \SI{450}{\pico \second} between the 1-photon and the 3-photon event.  
Because a fixed threshold is used, the 3-photon event with higher slew rate will reach the threshold much earlier than the 1-photon event. 
In contrast to that, CFD mode eliminates this effect almost completely, and the mean $t_\mathrm{ver}$ stays nearly constant and varies by less than \SI{110}{\pico \second}.
The remaining shift could be attributed to potentially faster pulse formation processes when all pixels fire, as well as reduced signal propagation length. 
In summary, we determined the individual photon-number-dependent jitter contributions and improved the operational stability of the CFD mode. 
This enables jitter characterization at the optimum threshold for all photon numbers simultaneously.
\\

\section{Summary and Conclusion}
To summarize, we started by showing the $n$-dependent output pulses of an SND. 
They confirm the linear increase of the pulse amplitude with $n$, while the time constants remain constant across $n$ within measurement uncertainty, yielding an average rise time of \SI{658}{\pico\second} and a fall time of \SI{14.3}{\nano\second}.
These results form the foundation of the dual-trigger method, which is applied to the waveforms, where we trigger on both the rising and falling edge of the output pulse. 
Separation of the signal into multiple channels allows careful adjustment of the individual timestamps. 
Ultimately, we show the 2D heatmap from the dual-trigger, where we plot the events based on their arrival time $t_\mathrm{ver}$ or \trise and their time over falling $\Delta t$. 
The resulting distribution is clustered into three islands, each island corresponding to a photon number $n$.
$t_\mathrm{ver}$ and \trise contain the information about the absolute arrival time, whereas the difference between the arrival time and \tfall serves as a separator for $n$. 
Moreover, we compare two trigger mechanisms on the rising edge, namely the CFD and edge-trigger. 
The main advantage of timestamps acquired in CFD mode is that they are determined by the vertex of the pulse, which eliminates the threshold dependence and compensates for pulse amplitude variations. \\ 

Further analyzing the data represented by the model we can assign photon numbers with a certainty of $99\,\%$ while having a leakage rate of unassigned photons of $0.01\,\%$. 
Most importantly, compared with previously shown techniques, once the boundaries are determined we can provide $n$ on an event-by-event basis rather than a probability distribution. 
In addition, it allows us to easily group the data by $n$ and extract the timing jitter for both measurement configurations.
The highest timing jitter of \SI{161}{\pico \second} is measured for the single-photon event using an edge-trigger. 
In contrast to that, the single-photon jitter with CFD mode is much smaller and results in \SI{60}{\pico \second}. 
For higher photon numbers, the timing jitter is smaller than $\SI{41}{\pico \second}$. 
To summarize, we compared the timing performance of using edge and CFD mode, yielding similar results for $n \geq 2$ and improved results for $n = 1$. \\

Taken together, the SND and a dual-trigger readout enable precise determination of the photon number on a single-shot basis.
It removes the need to split the signal into many channels for coincidence counting, which comes with the downside of requiring multiple coax lines and reduced amplitude and increased jitter due to the signal splitting. 
Moreover, this separation makes complex post-signal processing \cite{Hao2024} no longer necessary, allows for real-time signal processing, and enables the determination of $n$-dependent detector metrics. 
Besides timing jitter, it allows us to study many more detector parameters, such as dead time and efficiency. 
The CFD is identified as a useful tool for eliminating the effects of pulse amplitude variations and for further reducing the time bin size. 
We generally expect the CFD to be even more advantageous when repetition rates are higher than \SI{8}{\mega \hertz} as used in the shown experiments.
Further developments could include combining the presented readout mechanism with other circuit modifications. 
Different methods have been proposed to resolve the location of the detection event, either by variation of the shunt resistor value $R_\mathrm{p}$ \cite{He2022} or by adding various series inductors to the parallel path \cite{Guan2023}. 
Both methods vary the pulse shape, which would allow reconstruction of the photon's detection site. 
The question remains how much further the separation can be driven by detecting higher mean photon numbers and using more pixels. 
Therefore, next steps should include further tailoring of the detector circuit and the output pulse to ensure maximum spacing of the photon number encoding, which would be useful for many quantum applications. 

\section*{Acknowledgement}
We thank Martin Helversen, Marcel Hohn, Andreas Lehr und Christoph Tyborski  for the valuable insights and fruitful discussions and PicoQuant for providing measurement instrumentation. 
We also thank Christian Schmid and the team from Munich Quantum Instruments for providing the detector.
This work was supported from the German Federal Ministry of Research Technology and Space (BMFTR) via the funding program “Quantum technologies – from basic research to market” (projects SAEQS (3N16760), PhotonQ (13N15760), QPIS.2 (16KISQ172), FgED (13N17529), QPIS (16K1SQ033), QPIC-1 (13N15855), SPINNING (13N16214) and QR-N (16KIS2197)), as well as from the German Research Foundation (DFG) under Germany’s Excellence Strategy EXC-2111 (390814868) and projects INST 95/1720-1 (MQCL) and PQET (INST 95/1654-1). This research is also supported by the ‘Munich Quantum Valley”, which is supported by the Bavarian state government with funds from the “Hightech Agenda Bayern Plus”.

\section{Supplementary Material}
\subsection{Photon Level Dependent Pulse Properties}
\label{sec:pnr_pp}
\begin{figure}[h]
    \includegraphics[width=\columnwidth]{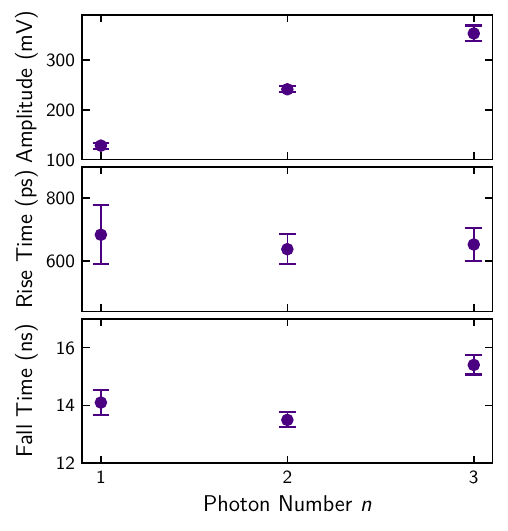}
    \caption{Pulse properties recorded with the oscilloscope. a) The pulse amplitude increases linearly with photon number. b) The rise time shows a mean value of \SI{658}{\pico \second} and is independent of photon number. c) The fall time is also independent of photon number with a mean value of \SI{14.3}{\nano \second}.}
    \label{fig:pnr_pp}
\end{figure}

To isolate a given photon number $n$ for characterization on an edge-triggered oscilloscope, we combine two complementary controls: the trigger threshold and the laser's mean photon number. 
The threshold is set above the pulse amplitude corresponding to $n-1$ photons, so that lower-order events do not trigger acquisition. 
The mean photon number is chosen, via attenuation, such that the probability of detecting more than $n$ photons is negligible. 
Together, these two conditions restrict the measured pulses to the $n$-photon level. 
\Cref{fig:pnr_pp} shows the average pulse properties of \SI{1500}{events} for the individual photon levels using a fast oscilloscope.
The standard deviation from the their mean value is displayed as an errorbar. 
The pulse height shows the maximum amplitude of the voltage transient. 
We note here that the pulse properties have been measured without the splitter. 
Therefore the amplitudes are a factor of four larger compared to \Cref{fig:PNR_waveforms}. 
As expected from the detector design, we observe a linear increase with photon number.
The 1-photon, 2-photon and 3-photon events have a mean pulse amplitude of \SI{128}{\milli \volt}, \SI{241}{\milli \volt} and \SI{353}{\milli \volt}, respectively.
The rise and fall times are defined as the time intervals between 90\% and 1/$e$ of the maximum amplitude. 
Both rise and fall times stay nearly constant with a mean of \SI{658}{\pico \second} and \SI{14.3}{\nano \second}, respectively. \\

\bibliography{references.bib}

@article{Applegate2015,
  title = {Efficient and Robust Quantum Random Number Generation by Photon Number Detection},
  author = {Applegate, M. J. and Thomas, O. and Dynes, J. F. and Yuan, Z. L. and Ritchie, D. A. and Shields, A. J.},
  year = 2015,
  month = aug,
  journal = {Applied Physics Letters},
  volume = {107},
  number = {7},
  pages = {071106},
  issn = {0003-6951},
  doi = {10.1063/1.4928732},
  urldate = {2026-06-21}
}

@article{Becerra2015,
  title = {Photon Number Resolution Enables Quantum Receiver for Realistic Coherent Optical Communications},
  author = {Becerra, F. E. and Fan, J. and Migdall, A.},
  year = 2015,
  month = jan,
  journal = {Nature Photonics},
  volume = {9},
  number = {1},
  pages = {48--53},
  issn = {1749-4885, 1749-4893},
  doi = {10.1038/nphoton.2014.280},
  urldate = {2026-06-21},
  langid = {english}
}

@article{Cahall2017,
  title = {Multi-Photon Detection Using a Conventional Superconducting Nanowire Single-Photon Detector},
  author = {Cahall, Clinton and Nicolich, Kathryn L. and Islam, Nurul T. and Lafyatis, Gregory P. and Miller, Aaron J. and Gauthier, Daniel J. and Kim, Jungsang},
  year = 2017,
  month = dec,
  journal = {Optica},
  volume = {4},
  number = {12},
  pages = {1534--1535},
  publisher = {Optica Publishing Group},
  issn = {2334-2536},
  doi = {10.1364/OPTICA.4.001534},
  urldate = {2024-01-06},
  copyright = {\copyright{} 2017 Optical Society of America},
  langid = {english}
}

@article{Dauler2007,
  title = {Multi-{{Element Superconducting Nanowire Single-Photon Detector}}},
  author = {Dauler, E.A. and Robinson, B.S. and Kerman, A.J. and Yang, J.K.W. and Rosfjord, E.K.M. and Anant, V. and Voronov, B. and Gol'tsman, G. and Berggren, K.K.},
  year = 2007,
  month = jun,
  journal = {IEEE Transactions on Applied Superconductivity},
  volume = {17},
  number = {2},
  pages = {279--284},
  issn = {1051-8223},
  doi = {10.1109/TASC.2007.897372},
  urldate = {2024-01-03},
  langid = {english}
}

@article{Fitch2003,
  title = {Photon-Number Resolution Using Time-Multiplexed Single-Photon Detectors},
  author = {Fitch, M. J. and Jacobs, B. C. and Pittman, T. B. and Franson, J. D.},
  year = 2003,
  month = oct,
  journal = {Physical Review A},
  volume = {68},
  number = {4},
  pages = {043814},
  publisher = {American Physical Society},
  doi = {10.1103/PhysRevA.68.043814},
  urldate = {2026-05-24}
}

@article{Flaschmann2023,
  title = {The Dependence of Timing Jitter of Superconducting Nanowire Single-Photon Detectors on the Multi-Layer Sample Design and Slew Rate},
  author = {Flaschmann, Rasmus and Zugliani, Lucio and Schmid, Christian and Spedicato, Simone and Strohauer, Stefan and Wietschorke, Fabian and Flassig, Fabian and J.~Finley, Jonathan and M{\"u}ller, Kai},
  year = 2023,
  journal = {Nanoscale},
  volume = {15},
  number = {3},
  pages = {1086--1091},
  publisher = {Royal Society of Chemistry},
  doi = {10.1039/D2NR04494C},
  urldate = {2024-04-29},
  langid = {english}
}

@article{Gouzien2018,
  title = {Quantum Description of Timing-Jitter for Single Photon {{ON}}/{{OFF}} Detectors},
  author = {Gouzien, {\'E}lie and Fedrici, Bruno and Zavatta, Alessandro and Tanzilli, S{\'e}bastien and D'Auria, Virginia},
  year = 2018,
  month = jul,
  journal = {Physical Review A},
  volume = {98},
  number = {1},
  eprint = {1807.00722},
  primaryclass = {quant-ph},
  pages = {013833},
  issn = {2469-9926, 2469-9934},
  doi = {10.1103/PhysRevA.98.013833},
  urldate = {2026-07-30},
  archiveprefix = {arXiv}
}

@article{Guan2023,
  title = {Approaching Pixel-Level Readout of {{SNSPD}} Array by Inductor-Shaping Pulse},
  author = {Guan, Yanqiu and Li, Haochen and Zhang, Labao and Dong, Daxing and Wang, Hao and Chen, Qi and Guo, Shuya and Zhang, Biao and Zhang, Xiao and Yang, Zhuolin and Tu, Xuecou and Zhao, Qingyuan and Jia, Xiaoqing and Chen, Jian and Kang, Lin and Wu, Peiheng},
  year = 2023,
  month = jul,
  journal = {Applied Physics Letters},
  volume = {123},
  number = {4},
  pages = {042602},
  issn = {0003-6951, 1077-3118},
  doi = {10.1063/5.0159725},
  urldate = {2024-07-06},
  langid = {english}
}

@article{Hao2024,
  title = {A Compact Multi-Pixel Superconducting Nanowire Single-Photon Detector Array Supporting Gigabit Space-to-Ground Communications},
  author = {Hao, Hao and Zhao, Qing-Yuan and Huang, Yang-Hui and Deng, Jie and Yang, Fan and Ru, Sai-Ying and Liu, Zhen and Wan, Chao and Liu, Hao and Li, Zhi-Jian and Wang, Hua-Bing and Tu, Xue-Cou and Zhang, La-Bao and Jia, Xiao-Qing and Wu, Xing-Long and Chen, Jian and Kang, Lin and Wu, Pei-Heng},
  year = 2024,
  month = jan,
  journal = {Light: Science \& Applications},
  volume = {13},
  number = {1},
  pages = {25},
  publisher = {Nature Publishing Group},
  issn = {2047-7538},
  doi = {10.1038/s41377-023-01374-1},
  urldate = {2024-04-02},
  copyright = {2024 The Author(s)},
  langid = {english}
}

@article{He2022,
  title = {Simultaneous Resolution of Photon Numbers and Positions with Series-Connected Superconducting Nanowires},
  author = {He, Guanglong and Li, Haochen and Yin, Rui and Zhang, Labao and Dong, Daxing and Lv, Jiayu and Fei, Yue and Wang, Xiaohan and Chen, Qi and Li, Feiyan and Li, Hui and Wang, Hao and Tu, Xuecou and Zhao, Qingyuan and Jia, Xiaoqing and Chen, Jian and Kang, Lin and Wu, Peiheng},
  year = 2022,
  month = mar,
  journal = {Applied Physics Letters},
  volume = {120},
  number = {12},
  pages = {124001},
  issn = {0003-6951, 1077-3118},
  doi = {10.1063/5.0084744},
  urldate = {2025-01-02},
  langid = {english}
}

@article{Huang2018,
  title = {High Speed Superconducting Nanowire Single-Photon Detector with Nine Interleaved Nanowires},
  author = {Huang, Jia and Zhang, Weijun and You, Lixing and Zhang, Chengjun and Lv, Chaolin and Wang, Yong and Liu, Xiaoyu and Li, Hao and Wang, Zhen},
  year = 2018,
  month = jul,
  journal = {Superconductor Science and Technology},
  volume = {31},
  number = {7},
  pages = {074001},
  issn = {0953-2048, 1361-6668},
  doi = {10.1088/1361-6668/aac180},
  urldate = {2023-11-02},
  langid = {english}
}

@article{Jahanmirinejad2012,
  title = {Proposal for a Superconducting Photon Number Resolving Detector with Large Dynamic Range},
  author = {Jahanmirinejad, Saeedeh and Fiore, Andrea},
  year = 2012,
  month = feb,
  journal = {Optics Express},
  volume = {20},
  number = {5},
  pages = {5017--5028},
  publisher = {Optica Publishing Group},
  issn = {1094-4087},
  doi = {10.1364/OE.20.005017},
  urldate = {2024-07-09},
  copyright = {\copyright{} 2012 OSA},
  langid = {english}
}

@article{Korzh2020,
  title = {Demonstration of Sub-3 Ps Temporal Resolution with a Superconducting Nanowire Single-Photon Detector},
  author = {Korzh, Boris and Zhao, Qing-Yuan and Allmaras, Jason P. and Frasca, Simone and Autry, Travis M. and Bersin, Eric A. and Beyer, Andrew D. and Briggs, Ryan M. and Bumble, Bruce and Colangelo, Marco and Crouch, Garrison M. and Dane, Andrew E. and Gerrits, Thomas and Lita, Adriana E. and Marsili, Francesco and Moody, Galan and Pe{\~n}a, Cristi{\'a}n and Ramirez, Edward and Rezac, Jake D. and Sinclair, Neil and Stevens, Martin J. and Velasco, Angel E. and Verma, Varun B. and Wollman, Emma E. and Xie, Si and Zhu, Di and Hale, Paul D. and Spiropulu, Maria and Silverman, Kevin L. and Mirin, Richard P. and Nam, Sae Woo and Kozorezov, Alexander G. and Shaw, Matthew D. and Berggren, Karl K.},
  year = 2020,
  month = apr,
  journal = {Nature Photonics},
  volume = {14},
  number = {4},
  pages = {250--255},
  publisher = {Nature Publishing Group},
  issn = {1749-4893},
  doi = {10.1038/s41566-020-0589-x},
  urldate = {2023-10-09},
  copyright = {2020 The Author(s), under exclusive licence to Springer Nature Limited},
  langid = {english}
}

@article{Los2024,
  title = {High-Performance Photon Number Resolving Detectors for 850--950~Nm Wavelength Range},
  author = {Los, J. W. Niels and Sidorova, Mariia and {Lopez-Rodriguez}, Bruno and Qualm, Patrick and Chang, Jin and Steinhauer, Stephan and Zwiller, Val and Zadeh, Iman Esmaeil},
  year = 2024,
  month = jun,
  journal = {APL Photonics},
  volume = {9},
  number = {6},
  pages = {066101},
  issn = {2378-0967},
  doi = {10.1063/5.0204340},
  urldate = {2026-05-24}
}

@article{Mueller2023,
  title = {Time-Walk and Jitter Correction in {{SNSPDs}} at High Count Rates},
  author = {Mueller, Andrew and Wollman, Emma E. and Korzh, Boris and Beyer, Andrew D. and Narvaez, Lautaro and Rogalin, Ryan and Spiropulu, Maria and Shaw, Matthew D.},
  year = 2023,
  month = jan,
  journal = {Applied Physics Letters},
  volume = {122},
  number = {4},
  pages = {044001},
  issn = {0003-6951, 1077-3118},
  doi = {10.1063/5.0129147},
  urldate = {2024-01-18},
  langid = {english}
}

@article{Nerenberg2025,
  title = {Photon Number-Resolving Quantum Reservoir Computing},
  author = {Nerenberg, Sam and Neill, Oliver D. and Marcucci, Giulia and Faccio, Daniele},
  year = 2025,
  month = apr,
  journal = {Optica Quantum},
  volume = {3},
  number = {2},
  pages = {201},
  issn = {2837-6714},
  doi = {10.1364/OPTICAQ.553294},
  urldate = {2026-07-29},
  langid = {english}
}

@article{OBrien2016,
  title = {Optical Quantum Computing},
  author = {O'Brien, Jeremy L.},
  year = 2016,
  journal = {Proceedings - International Conference on Natural Computation},
  volume = {2016-Janua},
  number = {December},
  pages = {390--397},
  issn = {21579555},
  doi = {10.1109/ICNC.2015.7378022},
  isbn = {9781467376792}
}

@article{Perrenoud2021,
  title = {Operation of Parallel {{SNSPDs}} at High Detection Rates},
  author = {Perrenoud, Matthieu and Caloz, Misael and Amri, Emna and Autebert, Claire and Sch{\"o}nenberger, Christian and Zbinden, Hugo and Bussi{\`e}res, F{\'e}lix},
  year = 2021,
  month = feb,
  journal = {Superconductor Science and Technology},
  volume = {34},
  number = {2},
  pages = {024002},
  issn = {0953-2048, 1361-6668},
  doi = {10.1088/1361-6668/abc8d0},
  urldate = {2024-07-06},
  langid = {english}
}

@article{Resta2023,
  title = {Gigahertz {{Detection Rates}} and {{Dynamic Photon-Number Resolution}} with {{Superconducting Nanowire Arrays}}},
  author = {Resta, Giovanni V. and Stasi, Lorenzo and Perrenoud, Matthieu and {El-Khoury}, Sylvain and Brydges, Tiff and Thew, Rob and Zbinden, Hugo and Bussi{\`e}res, F{\'e}lix},
  year = 2023,
  number = {23},
  pages = {6018--6026},
  doi = {10.1021/acs.nanolett.3c01228},
  urldate = {2023-10-11},
  langid = {english}
}

@article{Schapeler2024,
  title = {Electrical Trace Analysis of Superconducting Nanowire Photon-Number-Resolving Detectors},
  author = {Schapeler, Timon and Lamberty, Niklas and Hummel, Thomas and Schlue, Fabian and Stefszky, Michael and Brecht, Benjamin and Silberhorn, Christine and Bartley, Tim J.},
  year = 2024,
  month = jul,
  journal = {Physical Review Applied},
  volume = {22},
  number = {1},
  pages = {014024},
  issn = {2331-7019},
  doi = {10.1103/PhysRevApplied.22.014024},
  urldate = {2025-07-31},
  langid = {english}
}

@article{Schapeler2026a,
  title = {Practical Considerations for Assignment of Photon Numbers with {{SNSPDs}}},
  author = {Schapeler, Timon and Mischke, Isabell and Schlue, Fabian and Stefszky, Michael and Brecht, Benjamin and Silberhorn, Christine and Bartley, Tim J.},
  year = 2026,
  month = mar,
  journal = {APL Quantum},
  volume = {3},
  number = {1},
  eprint = {2510.00714},
  primaryclass = {quant-ph},
  pages = {016102},
  issn = {2835-0103},
  doi = {10.1063/5.0304127},
  urldate = {2026-07-14},
  archiveprefix = {arXiv}
}

@article{Sidorova2017,
  title = {Physical Mechanisms of Timing Jitter in Photon Detection by Current-Carrying Superconducting Nanowires},
  author = {Sidorova, Mariia and Semenov, Alexej and H{\"u}bers, Heinz-Wilhelm and Charaev, Ilya and Kuzmin, Artem and Doerner, Steffen and Siegel, Michael},
  year = 2017,
  month = nov,
  journal = {Physical Review B},
  volume = {96},
  number = {18},
  pages = {184504},
  issn = {2469-9950, 2469-9969},
  doi = {10.1103/PhysRevB.96.184504},
  urldate = {2024-05-05},
  copyright = {https://link.aps.org/licenses/aps-default-license},
  langid = {english}
}

@article{Sidorova2025,
  title = {Jitter in Photon-Number-Resolved Detection by Superconducting Nanowires},
  author = {Sidorova, Mariia and Schapeler, Timon and Semenov, Alexej D. and Schlue, Fabian and Stefszky, Michael and Brecht, Benjamin and Silberhorn, Christine and Bartley, Tim J.},
  year = 2025,
  month = aug,
  journal = {APL Photonics},
  volume = {10},
  number = {8},
  pages = {086113},
  issn = {2378-0967},
  doi = {10.1063/5.0273752},
  urldate = {2026-04-25}
}

@misc{Stasi2024,
  title = {High Photon-Number Efficiencies with a Fast 28-Pixel Parallel {{SNSPD}}},
  author = {Stasi, Lorenzo and Taher, Towsif and Resta, Giovanni V. and Zbinden, Hugo and Thew, Rob and Bussi{\`e}res, F{\'e}lix},
  year = 2024,
  month = jun,
  number = {arXiv:2406.15312},
  eprint = {2406.15312},
  primaryclass = {physics, physics:quant-ph},
  publisher = {arXiv},
  urldate = {2024-07-06},
  archiveprefix = {arXiv},
  langid = {english}
}

@article{Tao2020,
  title = {Characterize the {{Speed}} of a {{Photon-Number-Resolving Superconducting Nanowire Detector}}},
  author = {Tao, Xu and Hao, Hao and Li, Xiang and Chen, Shi and Wang, Libo and Tu, Xuecou and Jia, Xiaoqing and Zhang, Labao and Zhao, Qingyuan and Kang, Lin and Wu, Peiheng},
  year = 2020,
  month = aug,
  journal = {IEEE Photonics Journal},
  volume = {12},
  number = {4},
  pages = {1--8},
  issn = {1943-0655},
  doi = {10.1109/JPHOT.2020.3012349},
  urldate = {2024-07-09}
}

@article{Tiedau2020,
  title = {Single-Channel Electronic Readout of a Multipixel Superconducting Nanowire Single Photon Detector},
  author = {Tiedau, Johannes and Schapeler, Timon and Anant, Vikas and Fedder, Helmut and Silberhorn, Christine and Bartley, Tim J.},
  year = 2020,
  month = feb,
  journal = {Optics Express},
  volume = {28},
  number = {4},
  pages = {5528--5537},
  publisher = {Optica Publishing Group},
  issn = {1094-4087},
  doi = {10.1364/OE.383111},
  urldate = {2026-08-17},
  copyright = {\copyright{} 2020 Optical Society of America},
  langid = {english}
}

\end{document}